\documentclass[twocolumn,apl,amsmath,amssymb,showpacs]{revtex4-2}
\usepackage{epsf}      
\usepackage{graphicx}
\usepackage{color}
\usepackage{soul}
\usepackage{gensymb}
\usepackage{sidecap}
\usepackage{amsmath}
\usepackage{mathtools}
\usepackage[hidelinks,colorlinks=true,linkcolor=blue,citecolor=blue]{hyperref}

\begin{document}
\title{Case study of an exploratory high voltage NASICON-based Na$_4$NiCr(PO$_4$)$_3$ cathode material for sodium-ion batteries} 
\author{Madhav Sharma}
\affiliation{Department of Physics, Indian Institute of Technology Delhi, Hauz Khas, New Delhi-110016, India}
\author{Pooja Sindhu}
\affiliation{Department of Physics, Indian Institute of Technology Delhi, Hauz Khas, New Delhi-110016, India}
\author{Rajendra S. Dhaka}
\email{rsdhaka@physics.iitd.ac.in}
\affiliation{Department of Physics, Indian Institute of Technology Delhi, Hauz Khas, New Delhi-110016, India}

\date{\today}      

\begin{abstract}

We examine a new NASICON-type Na$_4$NiCr(PO$_4$)$_3$ material designed for high-voltage and multi-electron reactions for the sodium-ion batteries (SIBs). The Rietveld refinement of the X-ray diffraction pattern, using the R$\bar{3}$c space group, confirmed the stabilization of the rhombohedral NASICON framework. Furthermore, the Raman and Fourier transform infrared spectroscopy are employed to probe the structure and chemical bonding. The core-level photoemission analysis reveals the Cr$^{3+}$ and mixed Ni$^{2+}$/Ni$^{3+}$ oxidation states in the sample. Moreover, the bond valence energy landscape (BVEL) analysis, based on the refined structure, revealed a three-dimensional network of well-connected sodium sites with a migration energy barrier of 0.468 eV. The material delivered a good charge capacity at around 4.5 V, but showed no sodium-ion intercalation during discharge, resulting in negligible discharge capacity. The post-mortem analysis confirmed that the crystal structure remained intact. The calculated energy barrier values indicated a reversal in sodium site stability after cycling, though the barriers can still permit feasible ion migration. This suggests that ion transport alone cannot explain the lack of reversibility, which likely arises from intrinsically poor electronic conductivity. These findings highlight key challenges in achieving stable, reversible capacity in this system and underscore the need for doping, structural modification, and electrolyte optimization to realize its full potential as a high-voltage SIB cathode.

\end{abstract}

\maketitle
\section{\noindent~Introduction}

The growing demand for sustainable and large-scale energy storage technologies has intensified research into sodium-ion batteries (SIBs) as a cost-effective alternative of lithium-ion batteries (LIBs) \cite{Jin_CSR_20, Sapra_JMCA_25, Sapra_ACSAMI_24, Singh_CEJ_23}. Despite their promise, SIBs inherently operate at lower voltages due to the higher electrochemical potential of sodium compared to lithium, which limits their achievable energy density \cite{Sharma_CCR_25}. Addressing this challenge requires the development of high-voltage cathode materials. In this regard, transition metals such as Ni, Cr, and Mn are of particular interest because of their redox activity at elevated potentials \cite{Pati_JMCA_22, Kawai_CM_21, Sharma_S_25}. However, operation at high voltages can impose severe stresses on the battery system, leading to electrolyte decomposition, parasitic interfacial reactions, and irreversible structural transformations within the cathode. These degradation processes compromise long-term stability and reversibility \cite{You_AEM_18}. Consequently, attention has shifted toward cathode materials that not only offer high-voltage redox activity but also possess intrinsically robust frameworks capable of maintaining structural integrity under such demanding conditions \cite{Sapra_ESM_25} for rechargeable batteries. In the exploration of high-voltage cathodes, nickel remains in the focus owing to its redox activity at higher voltages, as a result modern high performance secondary LIBs are composed of LiNi$_{0.6}$Mn$_{0.2}$Co$_{0.2}$O$_2$ and LiNi$_{0.8}$Mn$_{0.1}$Co$_{0.1}$O$_2$ based formulations \cite{Li_NE_20, Wang_CSR_25}. In the case of SIBs, Ni-based layered oxides are regarded as highly promising candidates for achieving high-voltage operation, particularly when combined with Mn to balance cost and enhance the electrochemical performance \cite{Wang_JMCA_19}. However, these layered cathodes suffer from several critical challenges, including irreversible phase transitions, structural instability upon cycling, cation disorder, and pronounced capacity fading at high voltages \cite{Wang_JMCA_19, Goncalves_JMCA_24}. In order to address some of the issues, the polyanionic frameworks, particularly those based on the NASICON (NA Super Ionic CONductor) structure, provide a compelling platform owing to their three-dimensional Na$^+$ conduction channels and robust structural stability \cite{Meena_S_25, Sapra_WEE_21}. The open 3D framework with continuous ionic pathways ensures stable Na$^+$ insertion/extraction while mitigating the risk of structural collapse. Moreover, the NASICON cathodes have been reported to support multi-electron redox reactions with minimal volumetric changes, further enhancing their suitability for high-performance operation \cite{Liu_MF_23, Fan_MT_25}. These attributes make the NASICON framework an ideal host for exploring novel high-voltage cathode materials.

In this context, the Ni-based NASICON cathodes are relatively less explored in SIBs; for example, Zhou {\it et al.} attempted to synthesize Na$_4$NiV(PO$_4$)$_3$ cathode, but failed to obtain a pure phase \cite{Zhou_NL_16}. Kundu {\it et al.} tried the Na$_4$NiP$_2$O$_7$F$_2$ as a high-voltage cathode due to its predicted operating voltage of ~5 V. However, only about 0.35 Na$^+$ ions could be deintercalated during charging, and no significant electrochemical activity was observed during the subsequent intercalation process \cite{Kundu_CM_15}. The mixed ortho-pyro phosphate Na$_4$Ni$_3$(PO$_4$)$_2$P$_2$O$_7$ is an only known pure-phase polyanionic cathode capable of reversibly accommodating sodium ions, delivering a remarkable average working voltage of 4.8 V \cite{Zhang_NPGAM_17}. However, its discharge capacity was only 63 mAh/g, significantly lower than the theoretical value of 127.2 mAh/g, indicating the underperformance of the Ni$^{2+}$/Ni$^{3+}$ redox couple. The Na$_2$NiP$_2$O$_7$ was also explored electrochemically, but found inactive   \cite{Zhang_NPGAM_17}. Although, the Ni-based layered cathodes have established themselves as a class of high-voltage candidates, the performance of Ni redox couples in polyanionic frameworks has been far from satisfactory, and the reason could be the unavailability of the hole-polaron formation for the Ni in a polyanionic environment \cite{Johannes_PRB_12}. Moreover, the Na$_3$Cr$_2$(PO$_4$)$_3$ is attracting interest as a next-generation cathode due to its high Cr$^{3+}$/Cr$^{4+}$ redox voltage ($\sim$4.5 V), its theoretical capacity of 117 mAh/g, and an initial discharge capacity of 79 mAh/g, coupled with a low overpotential of $\approx$0.2 V. However, the discharge capacity degrades significantly within a few cycles owing to the uneasy nucleation of the discharged phase \cite{Kawai_ACSAEM_18}. Also, when half of the Cr is replaced with Ti, remarkable improvement in the cycling stability of the Cr$^{3+}$/Cr$^{4+}$ redox was observed \cite{Zhang_ACSAMI_20}. In case of NASICON cathodes, consisting of binary, ternary, or higher transition metal numbers, the inclusion of Cr is found to elevate the average working voltage \cite{Fan_MT_25, Zhang_ACSAMI_20, Wang_AEM_20}. 
 
Within this context, the NASICON materials, especially those incorporating redox-active elements such as Ni and Cr, present less explored yet promising route toward realizing SIBs with high-energy density. Considering their high-voltage activity, we selected the Na$_4$NiCr(PO$_4$)$_3$ (NNCP) binary system, stabilized within the flexible NASICON framework, as a potential candidate for high-voltage cathodes. The NNCP was first synthesized by Manoun {\it et al.}, who reported its stabilization in the rhombohedral NASICON phase \cite{Manoun_PD_04}. Subsequently, no further experimental studies were reported until Singh {\it et al.} conducted theoretical calculations, predicting redox activity of the Cr$^{3+}$/Cr$^{4+}$, Ni$^{2+}$/Ni$^{3+}$, and Ni$^{3+}$/Ni$^{4+}$ couples at 4.0, 4.3, and 4.4 V, respectively \cite{Singh_JMCA_21}. Being sodium-rich, the NNCP could accommodate up to three Na$^+$ ions per formula unit, with a predicted average voltage of 4.21 V and a theoretical capacity of 164.94 mAh/g, corresponding to an exceptional theoretical energy density of 694.40 Wh/kg. These promising figures motivate the electrochemical exploration of this material to advance the development of high-voltage cathodes for SIBs. Therefore, 
in this study, we successfully synthesized NNCP cathode material and characterized its structure using X-ray diffraction, Raman and Fourier transform infrared spectroscopy (FTIR). The BVEL analysis confirmed the presence of interconnected Na$^+$ transport pathways. However, the DC polarization measurements indicated low electronic conductivity, inherent in the polyanionic frameworks. To evaluate its electrochemical performance, we examined the redox activity, charge-discharge behavior, and reversibility of sodium storage in NNCP cathode. 

\section{\noindent ~Experimental}

{\bf{Synthesis of cathode material:}}
The NASICON-based NNCP polyanionic was synthesized using the sol-gel method using stoichiometric amounts of sodium nitrate, nickel nitrate hexahydrate, chromium nitrate nonahydrate, and ammonium dihydrogen phosphate as precursors. Three distinct solutions were prepared: Sol. I comprised 6 mmol of nickel nitrate hexahydrate and 6 mmol of chromium nitrate nonahydrate dissolved in 100 mL of deionized water. Sol. II involved dissolving 24 mmol of sodium nitrate in 20 mL of deionized water. Sol. III was prepared by dissolving 18 mmol of ammonium dihydrogen phosphate in 40 mL of deionized water. The Sol. II was added drop wise to Sol. I and stirred for an hr. Subsequently, Sol. III was slowly added to the combined solution. The resulting mixture was stirred for an additional hr before being heated to 80$\degree$C until a gel formed. This gel was then vacuum-heated overnight at 100$\degree$C to remove excess moisture. The resulting precursor was thoroughly ground for two hrs and subsequently subjected to heat treatment at 400$\degree$C, 600$\degree$C, and 750$\degree$C, each for 12 hrs in ambient atmosphere with intermittent grinding. The final product obtained under these conditions was labeled as NNCP-Air. For comparison, the same precursors were sintered under an argon atmosphere, first at 400$\degree$C for 6 hrs and then at 750$\degree$C for 12 hrs, yielding the sample denoted as NNCP-Ar. In addition, an {\it ex-situ} carbon coating was performed by mechanically mixing the NNCP-Air sample with 10 wt\% acetylene black, followed by heat treatment at 600$\degree$C for 6 hrs, and the resulting composite was designated as NNCP-Air/AB.

\begin{figure*}
\includegraphics[width=7.1in]{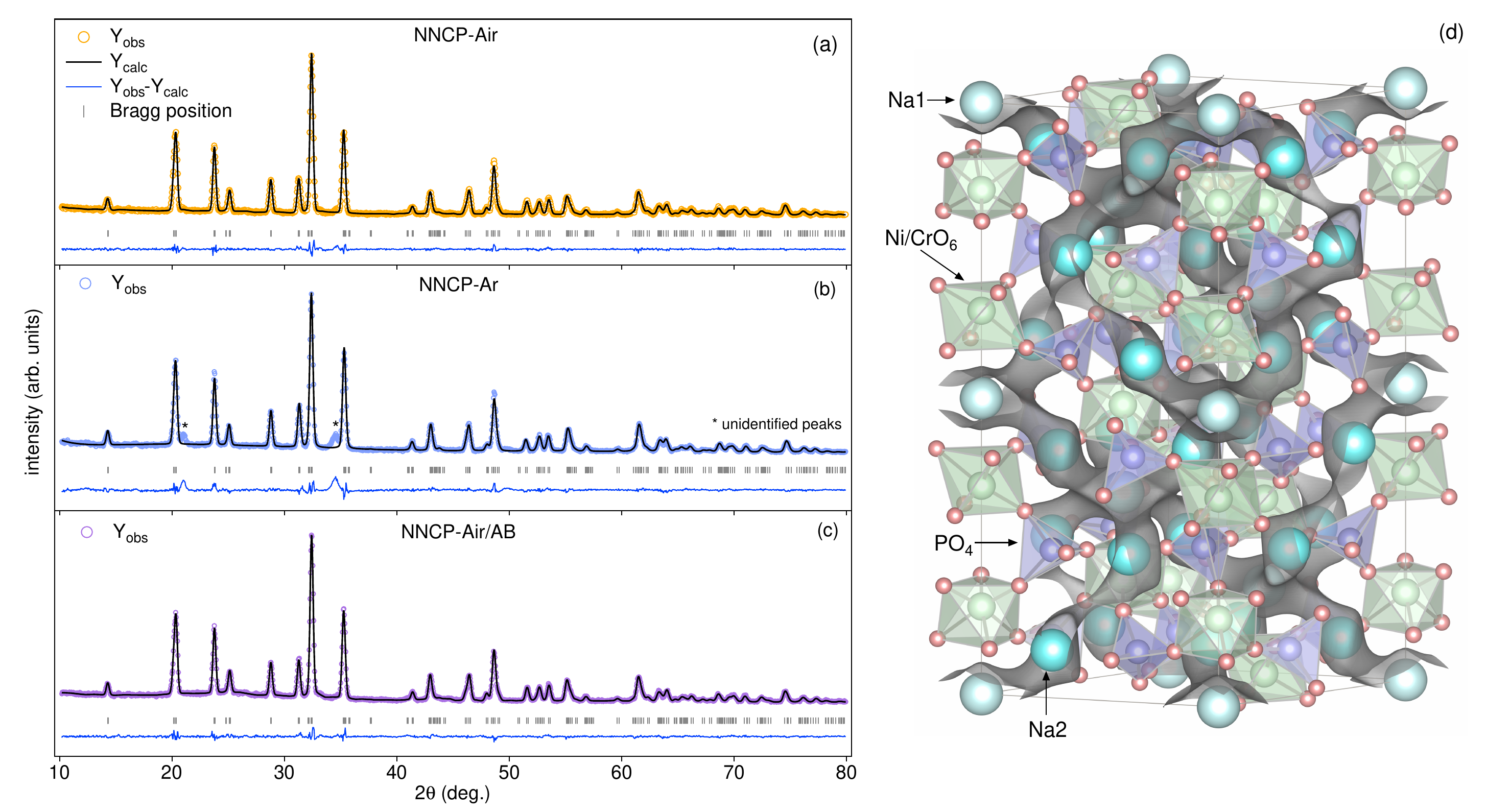}
\caption {Rietveld-refined XRD patterns of (a) NNCP-Air, (b) NNCP-Ar, and (c) NNCP-Air/AB samples, and (d) crystal structure of NNCP-Air showing calculated iso-surfaces (in grey colour) for 3D Na$^+$ migration channels.} 
\label{Strc}
\end{figure*}

{\bf{Physical characterization:}}

X-ray diffraction was performed utilizing a Panalytical Xpert$^3$ device at standard room temperature, utilizing Cu-K$\alpha$ radiation with a wavelength of 1.5406 \AA. The Raman spectrum was obtained using a Renishaw inVia confocal microscope outfitted with a 532 nm wavelength laser and a grating with 2400 lines per mm. The FTIR measurements were performed on a Thermo Nicolet IS50 instrument. The powdered samples were mixed with KBr and pressed into pellets prior to analysis. The field emission scanning electron microscopy (FESEM) and energy-dispersive X-ray spectroscopy (EDX) were performed on the sample using JEOL-FESEM (EVO18). The DC polarization measurement was taken using a Keithley 2450 Source Meter. The X-ray photoelectron spectroscopy (XPS) measurements were conducted using the Kratos Analytical Ltd spectrometer (AXIS SUPRA model) equipped with a monochromatic Al K$\alpha$ source (h$\nu$ = 1486.6 eV). The core-level spectra were acquired with a step size of 0.1 eV and 20 eV pass energy.

{\bf{Half-cell fabrication:}}

Electrochemical testing was performed using 2032-type half cells, assembled with NNCP as the working electrode, sodium metal as the counter/reference electrode, and 1 M NaPF$_6$ in EC:DEC (1:1 v/v) as the electrolyte. The electrode slurry was prepared by mixing NNCP, Super P, and PVDF in a 70:20:10 ratio using N-methylpyrrolidone (NMP) as the solvent. The slurry was cast onto aluminum foil by the doctor blade method, dried overnight at 80$\degree$C, and then calendared before punching 12 mm diameter electrodes. The coin-cell assembly was prepared  in an argon-filled glove box (UniLab Pro SP, MBraun) with O$_2$ and H$_2$O levels below 0.1 ppm.

{\bf{Electrochemical characterization:}}

The electrochemical performance is tested through detailed analysis of electrochemical impedance spectroscopy (EIS) and cyclic voltammetry (CV), which were performed using the Biologic VMP-3 model. The EIS covered a frequency range from 10 mHz to 100 kHz, with an alternating current voltage amplitude of 10 mV applied at the cell's OCV state. The galvanostatic charge-discharge (GCD) profiles were obtained using a Neware battery analyzer, BTS400.

\section{\noindent ~Results and discussion}

The long-range crystal structure and phase purity are analyzed by Rietveld refinement of the powder X-ray diffraction (XRD) pattern of the NNCP-Air sample, as shown in Fig.~\ref{Strc}(a). The calculated pattern shows excellent agreement with the rhombohedral NASICON phase (space group R$\bar{3}$c). For the refinement, the crystal structure of Na$_3$V$_2$(PO$_4$)$_3$ is adopted as the initial model, and the precise overlap between the calculated Bragg positions and the observed diffraction peaks confirms the accuracy of the structural model \cite{Sharma_IJPAP_24}. The refinement yielded lattice parameters of $a$ = $b$ = 8.8165 \AA~ and $c$ = 21.2658 \AA, which are in good agreement with the reported values for NNCP \cite{Manoun_PD_04}. The obtained atomic coordinates and site occupancies are summarized in Table \ref{tab:XRD}. The NNCP cathode is also synthesized under an inert argon atmosphere, referred to as NNCP-Ar. The Rietveld-refined XRD pattern [Fig. \ref{Strc}(b)] shows that most diffraction peaks closely match those of the sample synthesized in air and the calculated pattern, with a few minor unidentified impurity peaks. In contrast, the NNCP-Air/AB sample exhibits a similar XRD pattern [Fig. \ref{Strc}(c)] that aligns well with the calculated pattern, showing no additional unidentified peaks. To evaluate the possibility for ionic mobility, a BVEL analysis was performed \cite{Chen_ACB_19}. The resulting isosurface (shown in grey) highlights the low-energy pathways available for Na$^+$ diffusion, as depicted in Fig. \ref{Strc}(d). The analysis reveals a continuous, interconnected network connecting the Na1 and Na2 sites, forming a three-dimensional spread pathway throughout the structure with a sodium-ion migration energy barrier of 0.468 eV. This feature strongly suggests facile sodium-ion transport and emphasizes the material’s potential as an efficient sodium-ion conductor \cite{Sharma_S_25, Chen_ACB_19}. However, it should be noted that favorable ionic conduction alone does not guarantee good electrochemical performance; adequate electronic conductivity is equally essential to ensure efficient charge transfer during the redox processes of the active material \cite{Sharma_CCR_25}. 

\begin{table}[]
\centering
\caption{Structural and atomic parameters obtained from the Rietveld refinement of the XRD pattern of NNCP-Air.}
\label{tab:XRD}
\begin{tabular}{p{0.7cm}p{1.4cm}p{1.4cm}p{1.6cm}p{1.6cm}p{0.8cm}}
\hline
\multicolumn{3}{l}{space group = R$\bar{3}$c}                                & \multicolumn{3}{l}{$\chi^2$ = 1.54}                                         \\
\multicolumn{2}{l}{a(\AA) = 8.82} & \multicolumn{2}{l}{c(\AA) = 21.27} & \multicolumn{2}{l}{V(\AA) = 1431.54} \\ \hline
atom                    & x                      & y                         & z                     & occ.                     & site                     \\ \hline
Na1                   & 0.0000                 & 0.0000                    & 0.0000                & 1.000                    & 6b                       \\
Na2                   & 0.6395                 & 0                         & 0.2500                & 0.928                    & 18e                      \\
Ni                      & 0.0000                 & 0.0000                    & 0.1470                & 0.488                    & 12e                      \\
Cr                      & 0.0000                 & 0.0000                    & 0.1470                & 0.472                    & 12e                      \\
P                       & 0.2939                 & 0                         & 0.2500                & 0.960                    & 18e                      \\
O1                    & 0.1889                 & 0.1714                    & 0.0851                & 0.951                    & 36f                      \\
O2                    & 0.0195                 & 0.2041                    & 0.1910                & 0.893                    & 36f                      \\ \hline
\multicolumn{3}{l}{bond length (\AA)}                         & \multicolumn{3}{l}{bond angle (\degree)}                      \\ \hline
\multicolumn{2}{l}{Ni/Cr--O1}                   & 2.07                      & \multicolumn{2}{l}{$\angle$ Ni/Cr--O1--P}        & 141.85                   \\
\multicolumn{2}{l}{Ni/Cr--O2}                    & 1.96                      & \multicolumn{2}{l}{$\angle$ Ni/Cr--O2--P}        & 151.81                   \\
\multicolumn{2}{l}{P--O1}                        & 1.53                      & \multicolumn{2}{l}{$\angle$ O1--P--O2}           & 113.26                   \\
\multicolumn{2}{l}{P--O2}                        & 1.54                      & \multicolumn{2}{l}{$\angle$ Ni/Cr--Ni/Cr--Ni/Cr} & 80.68                    \\
\multicolumn{2}{l}{Ni/Cr--P}                     & 3.39                      & \multicolumn{2}{l}{$\angle$ Ni/Cr--P--Ni/Cr}     & 80.41                    \\
\multicolumn{2}{l}{O1--O2}                       & 2.57                      & \multicolumn{2}{l}{$\angle$ Ni/Cr--O2--Ni/Cr}    & 99.99   \\ \hline                
\end{tabular}
\end{table}

\begin{figure*}
	\includegraphics[width=7.2in]{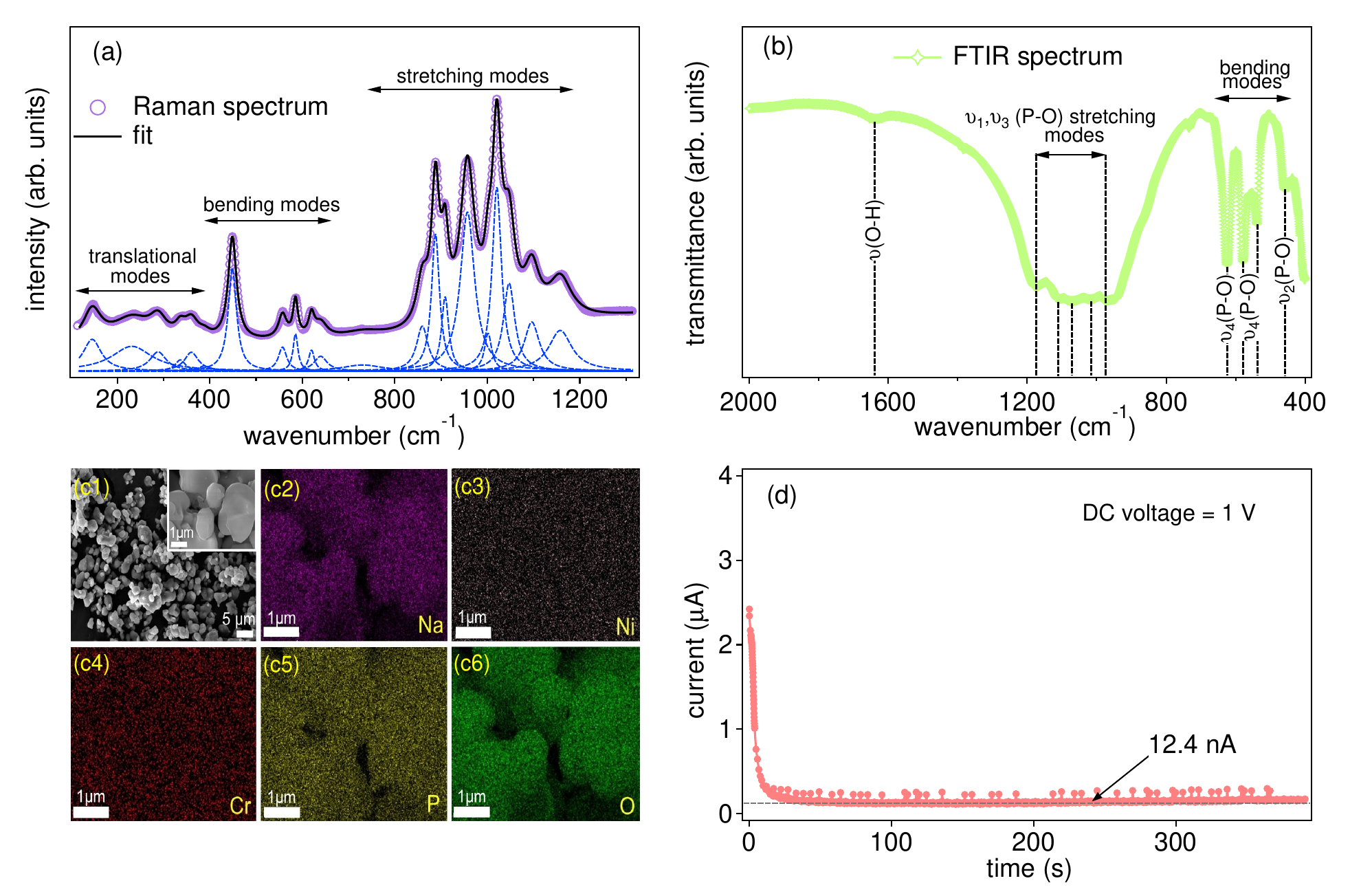}
	\caption{(a) The Raman, (b) FTIR spectra, (c1) the FE-SEM image, (c2-c6) the EDS elemental mappings, and (d) the DC polarization curve at a voltage of 1 V of the NNCP-Air sample.}
	\label{DC}
\end{figure*}

The Raman and FTIR spectroscopy are further employed to probe the local vibrational modes of the polyanionic framework, as shown in Figs.~\ref{DC}(a) and (b), respectively. The Raman spectrum display distinct features that can be categorized into three regions. The high-wavenumber bands, between 800-1200 cm$^{-1}$, arise from symmetric and asymmetric P--O stretching vibrations of the PO$_4$ tetrahedra \cite{Popovic_JRS_05}. The intermediate region from 400-800 cm$^{-1}$ corresponds to O--P--O bending modes of the phosphate groups along with contributions from NiO$_6$ octahedral vibrations. The bands observed in the low-wavenumber region, particularly between 300 and 400 cm$^{-1}$, are attributed to Cr--O stretching vibrations \cite{Barj_JSSC_92}. The low-wavenumber region below 300 cm$^{-1}$ is assigned to external lattice modes, including translational motions of the polyanionic units together with the Ni/CrO$_6$ octahedral units \cite{Singh_PRB_24}. In condensed phosphate systems, the P--O stretching vibrations show a strong correlation with the corresponding bond lengths. The relationship between the fundamental (average) wavenumber of the P--O stretching modes and the P--O bond length ($R$, in pm) is as below \cite{Popovic_JRS_05}:
\begin{equation}
\label{popovic}
    \nu_\omega = 224500\cdot\exp\left(\frac{-R}{28.35}\right)
\end{equation}
The crystallographically determined P--O bond lengths listed in Table \ref{tab:XRD} are substituted into the empirical relation equation~\ref{popovic}, yielding estimated fundamental wavenumbers of 1002.13 and 988.13 cm$^{-1}$ for the P--O1 and P--O2 bonds, respectively, with an average value of 995.11 cm$^{-1}$. These calculated values are in close agreement with the experimentally observed stretching modes in the range of 800--1200 cm$^{-1}$ [see Fig. \ref{DC}(a)]. This correspondence further confirms that the P--O bond lengths and their Raman stretching frequencies are exponentially related \cite{Popovic_JRS_05}. The FTIR spectrum of the sample, shown in Fig. \ref{DC}(b), exhibits prominent absorption bands within the 400-1200 cm$^{-1}$ region, which primarily correspond to the characteristic stretching and bending vibrations associated with the PO$_4$ tetrahedral units. The vibrational mode around 458 cm$^{-1}$ corresponds to symmetric bending mode ($\nu_2$) while the modes at 538 cm$^{-1}$ and 590 cm$^{-1}$ corresponds to asymmetric bending modes ($\nu_4$) of PO$_4$. In the higher wavenumber region, the peak near 965 cm$^{-1}$ is assigned to the symmetric stretching vibrations ($\nu_1$) of the PO$_4$ tetrahedra, while the bands appearing in the range of 1000-1200 cm$^{-1}$ correspond to asymmetric stretching modes ($\nu_3$) \cite{Sapra_ACSAMI_24}. 
The morphology and the elemental composition of the sample were further examined using FE-SEM and EDX. The FE-SEM image [Fig. \ref{DC}(c1)] depicted that the particles exhibit an irregular morphology and are agglomerated. The EDS mapping, as shown in Figs.~\ref{DC}(c2-c6), of all the elements Na, Ni, Cr, P, and O confirms the homogeneous distribution of the constituent elements across the analyzed area shown in the inset of Fig. \ref{DC}(c1). The zoomed FE-SEM image and the EDS mass sum spectrum are displayed in the Fig.~S1(a, b) of \cite{SI}. The electronic conductivity is evaluated using DC polarization, distinguishing ionic from electronic conductivity by applying a 1 V potential using Na-ion blocking Ag contacts \cite{Deivanayagam_BS_21}. Figure~\ref{DC}(d) depicts that the initial transient currents, which reflect both ionic and electronic conduction, decayed rapidly within a few seconds and reached a steady state after $\sim$50 s, attributed to electronic conduction, from which the calculated electronic conductivity is found to be 6.57$\times$10$^{-9}$ S/cm \cite{Choi_ACSAEM_24}. This extremely low value is likely associated with the unstable polaronic state of Ni in the NNCP framework, causing poor electronic conduction \cite{Johannes_PRB_12}.

\begin{figure}[h]
\includegraphics[width=3.4in]{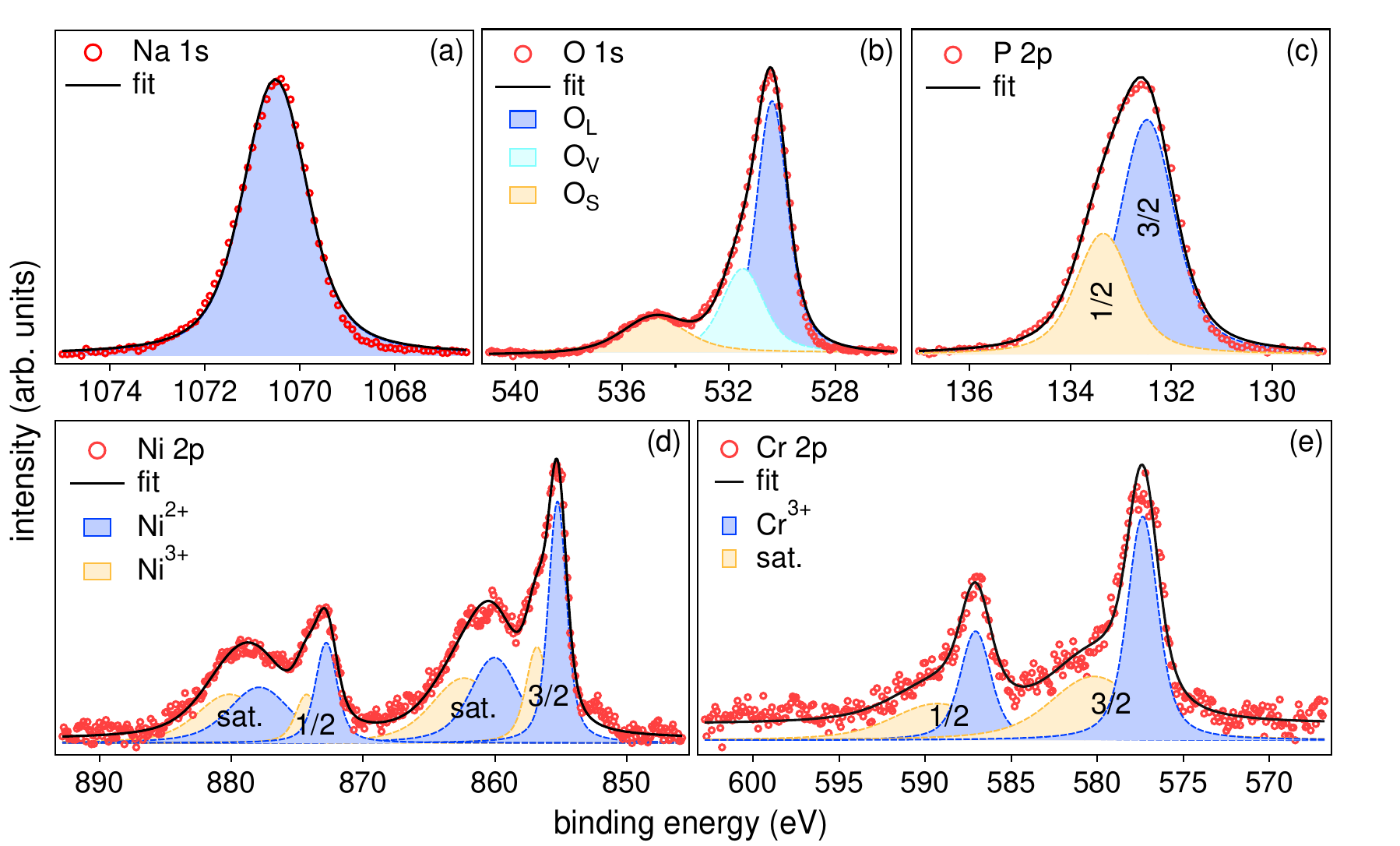}
\caption {The core level photoemission spectra of (a) Na 1$s$, (b) O 1$s$ ,(c) P 2$p$, (d) Ni 2$p$, and (e) Cr 2$p$.}
\label{XPS}
\end{figure}

Further, we use X-ray photoemission spectroscopy to investigate the valence states and chemical environment of different elements in the NNCP sample, and the corresponding core level spectra are presented in Figs.~\ref{XPS}(a-e). The survey spectrum shown in Fig.~S2 of \cite{SI} verifies the presence of all constituent elements. The binding energy of each spectrum,  shown in Figs. \ref{XPS}(a-e), is calibrated using the C 1s peak at 284.6 eV. After applying Tougaard background subtraction of the inelastic background, the core-level features were fitted in IGOR PRO software using Voigt profiles with a shape factor of 0.6. Figure~\ref{XPS}(a) shows the Na 1$s$ core level centered at 1070.5 eV, which represents the monovalent state of sodium \cite{PH_PRB_73}. The O 1$s$ core-level, presented in Fig.~\ref{XPS}(b) displays an intense peak around 530.4 eV (O$_L$), which is attributed to the lattice oxygen with a 2- formal charge, while the one with the lowest intensity on the higher binding energy side around 534.8 eV (O$_S$) corresponds to the surface contamination, i.e., hydroxylation of the surface species and the peak corresponding to 531.5 eV (O$_V$) is attributed to oxygen vacancies \cite{PH_PRB_73, AK_JAP_20}. As shown in Fig.~\ref{XPS}(c), the P 2$p$ core-level spectrum is deconvoluted with the spin-orbit coupled 2p$_{3/2}$ and 2p$_{1/2}$ components, separated by approximately 0.9 eV. The 2p$_{3/2}$ and 2p$_{1/2}$ features are centered at 132.5 and 133.4 eV, respectively \cite{LC_JPC_10}, indicating P in 5+ state. The deconvoluted Ni 2$p$ spectrum [see Fig.~\ref{XPS}(d)] consists of the characteristic spin-orbit doublet at 855.2 eV (2p$_{3/2}$) and 872.8 eV (2p$_{1/2}$), corresponding to Ni$^{2+}$. In addition, the features at 858.8 eV (2p$_{3/2}$) and 874.3 eV (2p$_{1/2}$) are assigned to Ni$^{3+}$ \cite{YY_AC_18}. Also, broad satellite features are observed at higher binding energies. The peaks at 860.1 eV and 877.9 eV represent the satellites of spin-orbit split components of Ni$^{2+}$, while the peaks at 862.3 eV and 880.1 eV originate from Ni$^{3+}$. The Cr 2$p$ region [Fig.~\ref{XPS}(e)] displays a well-defined doublet at 577.4 eV (2p$_{3/2}$) and 587.1 eV (2p$_{1/2}$), confirming the presence of Cr in the 3+ oxidation state \cite{JL_CEJ_21}. To maintain the overall charge neutrality in the presence of Ni$^{3+}$ ions, the charge compensation is achieved through the formation of oxygen vacancies in the lattice. These oxygen deficiencies effectively balance the excess positive charge arising from the partial oxidation of Ni$^{2+}$ to Ni$^{3+}$ in the pristine sample. 

\begin{figure*}
\includegraphics[width=7.2in]{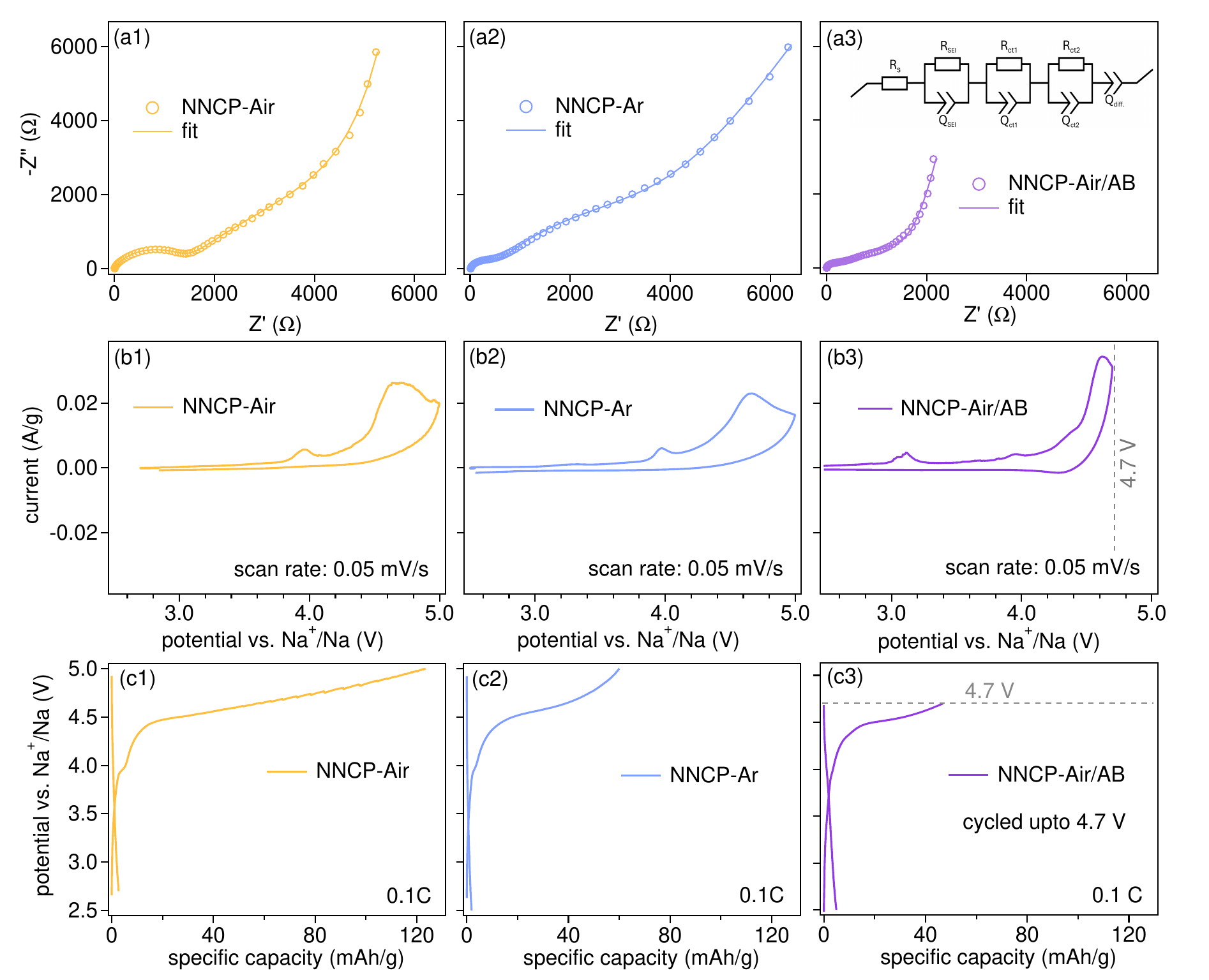}
\caption {(a1-a3) The EIS spectra with the equivalent circuit shown in the inset of (a3), (b1-b3) the cyclic voltammetry curves recorded at 0.05 mV/s, and (c1-c3) the galvanostatic charge-discharge profiles measured at 0.1 C for NNCP-Air, NNCP-Ar, and NNCP-Air/AB composite samples, respectively.}
\label{CV_GCD}
\end{figure*}

The electrochemical impedance spectra (EIS) were obtained on freshly fabricated cells and fitted with a circuit model consisting of a series resistance (R$_s$) followed by three RQ elements and an additional Q element (Q; constant phase element). The R$_s$ corresponds to the bulk resistance of the electrolyte and cell components. The first RQ element represents the resistance and capacitance of surface films at the electrode-electrolyte interface and is labeled as R$_{SEI}$. The second RQ element is attributed to intermediate interfacial processes, such as grain boundary resistance or surface reconstruction within the porous electrode. It can also be considered as a secondary charge transfer resistance and is labeled as R$_{ct1}$. The third RQ element reflects the primary charge-transfer resistance R$_{ct2}$ associated with Na$^+$ intercalation/deintercalation. Finally, the additional constant phase element captures the low-frequency diffusion-related processes within the electrode material \cite{Sapra_ESM_25, Mudgal_PINSA_22}. The EIS spectra of the fresh cells prepared using NNCP-Air and NNCP-Ar are shown in Figs.~\ref{CV_GCD}(a1,a2), both exhibiting relatively high overall impedance. From the equivalent circuit [shown in inset of Fig. \ref{CV_GCD}(a3)] fitting, the extracted resistances are as follows: R$_{SEI}$ = 172.8 $\Omega$ and 313.4 $\Omega$, and R$_{ct2}$ = 3518 $\Omega$ and 2998 $\Omega$ for NNCP-Air and NNCP-Ar, respectively. While these values are comparable between the two samples, a notable contrast emerges in R$_{ct1}$. This resistance drops markedly from 1167 $\Omega$ in NNCP-Air to 322.3 $\Omega$ in NNCP-Ar, highlighting a substantial improvement in secondary charge‐transfer kinetics for NNCP-Ar electrode.

\begin{figure*}
\includegraphics[width=7.2in]{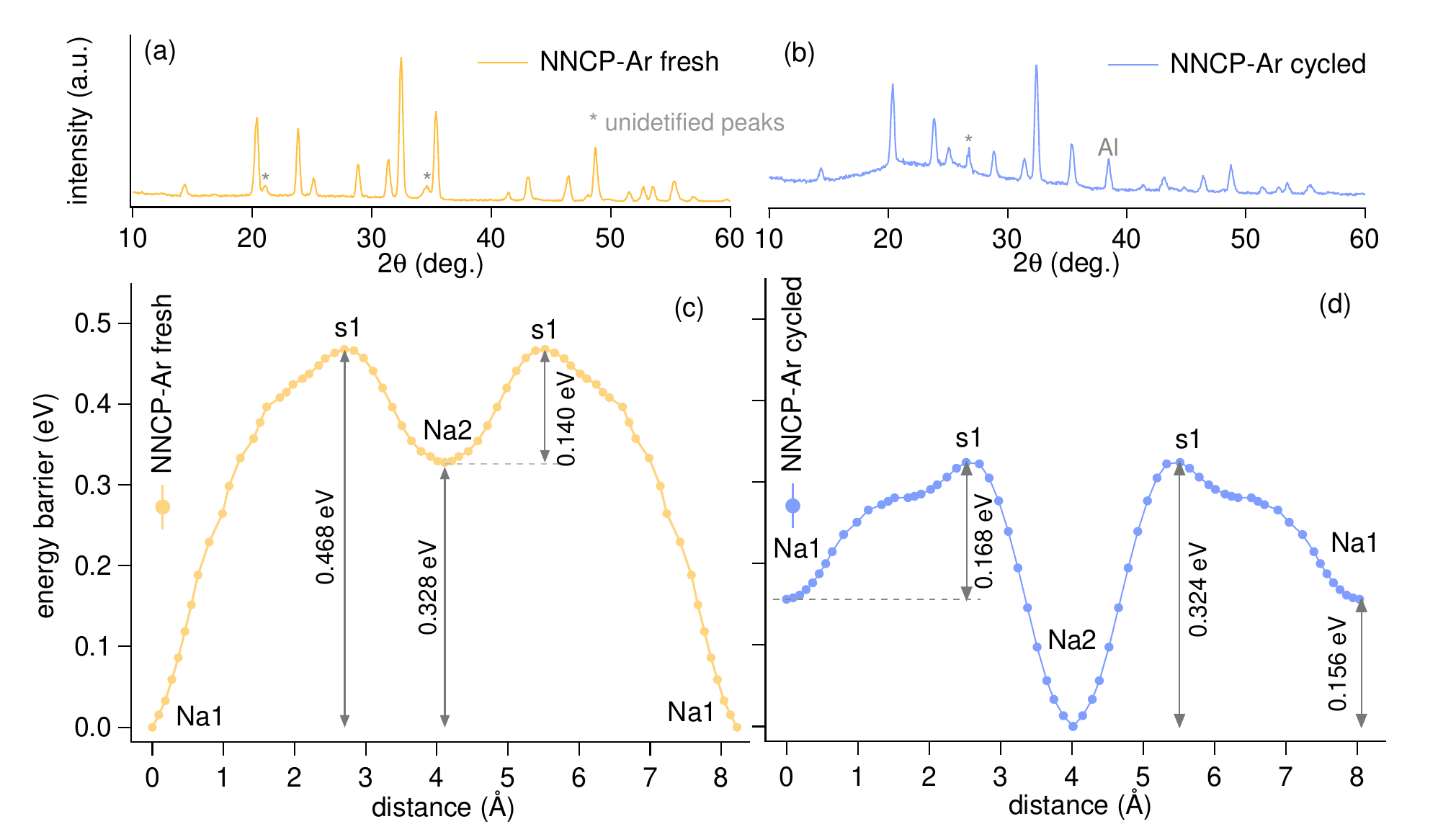}
\caption {The XRD pattern of (a) NNCP-Ar pristine sample and (b) NNCP-Ar electrode after cycling. A reaction pathway showing the 3D pathway network, in the (c) NNCP-Ar fresh sample and (d) NNCP cycled cathode, based on hopping between Na1 and Na2 sites.}
\label{Cycled}
\end{figure*}

The Cyclic voltammetry (CV) measurements are carried out to probe the redox behavior of the NNCP-Air sample [see Fig. \ref{CV_GCD}(b1)] in a potential window of 2.7--5.0 V. During the anodic scan, a small peak is observed at 3.9 V, followed by a broad and intense peak centered around 4.5 V. The latter aligns well with the electrochemical activity of the Cr$^{3+}$/Cr$^{4+}$ couple, which has been experimentally reported near 4.5 V in NASICON-type phosphates \cite{Kawai_ACSAEM_18, Zhang_ACSAMI_20}. We therefore attribute this feature to the oxidation of Cr$^{3+}$ to Cr$^{4+}$. However, during the subsequent reduction scan, no corresponding peaks is observed, as also shown in Fig.~S3(a) of \cite{SI} up to 3 cycles, which highlights the irreversible nature of the electrochemical processes. Similarly, the galvanostatic charge-discharge measurements are carried out, yielding a charge capacity of 123.5 mAh/g, as depicted in Fig.~\ref{CV_GCD}(c1). The profile revealed a small plateau at 3.9 V and a broad sloping plateau around 4.5 V, consistent with the features observed in the CV curves. However, during discharge, no evidence of reversible sodium intercalation is observed, and the cell delivered only a negligible discharge capacity of 2.7 mAh/g, which persisted in subsequent cycles [Fig. S3(b) of \cite{SI}], confirming the irreversible electrochemical behavior of this cathode. Note that the electrochemical measurements performed are restricted below the oxidative stability limit of 5 V for 1 M NaPF$_6$ in EC:DEC (1:1 v/v) \cite{Bhide_PCCP_14}. In contrast, severe electrolyte degradation is observed above 4.7 V for 1 M NaClO$_4$ in EC:DEC (1:1 v/v), as depicted in Fig.~S4 of \cite{SI}, and therefore, this electrolyte was not considered for further electrochemical testing \cite{Bhide_PCCP_14}. The absence of the 3.9 V plateau in the NaClO$_4$ electrolyte (Fig.~S4 of \cite{SI}) hints that the feature observed in NaPF$_6$ could potentially stem from the choice of electrolyte, rather than clear transition-metal redox activity. Furthermore, the EIS spectra of the NNCP-Air sample, recorded after five GCD cycles and compared with the fresh cell (Fig.~S5, \cite{SI}), showed a pronounced increase in impedance and a significant change in the Nyquist profile. These changes indicate severe deterioration of interfacial charge-transfer kinetics, consistent with the irreversible electrochemical behavior observed during cycling.

Additionally, the NNCP-Ar sample prepared under an argon atmosphere is also subjected to electrochemical testing in a slightly wider voltage window of 2.5-5.0 V, and the overall results are comparable to those of the NNCP-Air cathode. The CV curves of NNCP-Ar appeared smoother, without the fluctuations [see Fig. \ref{CV_GCD}(b2)], as observed in NNCP-Air, but similarly showed no signatures of sodiation in the subsequent cycles, as depicted in Fig.~S6(a) of \cite{SI}. The GCD profile shown in Fig.~\ref{CV_GCD}(c2) likewise exhibited similar features, with a more distinct plateau region compared to the sloping region near the cut-off voltage. The charge capacity in this case reached 60 mAh/g, nearly half of that obtained for NNCP-Air. Despite the smoother CV and GCD profiles, the discharge capacity remained unchanged at $\sim$2 mAh/g, indicating no improvement in reversibility [see Fig.~S6(b) of \cite{SI}]. Note that in the pristine state, the transition metals exist predominantly as Cr$^{3+}$ and Ni$^{2+}$. Upon charging, the Cr$^{3+}$/Cr$^{4+}$ redox couple is expected to contribute to the charge capacity and act as the primary pathway for polaronic electronic conduction \cite{Johannes_PRB_12, Kawai_ACSAEM_18}. During discharge, the Cr$^{4+}$ should be reduced back to Cr$^{3+}$ accompanied by Na$^+$ reinsertion; however, earlier reports have indicated an unfavorable nucleation of the discharged Cr$^{3+}$-containing phase \cite{Kawai_ACSAEM_18}. Consequently, the absence of hole-polaron formation associated with Ni and the ineffective re-establishment of the Cr$^{3+}$/Cr$^{4+}$ redox pairs severely limit the already poor electronic conductivity of the system, leading to a substantial increase in charge-transfer resistance that ultimately restricts sodium-ion re-insertion. 

Notably, the surface carbon coating is a widely adopted strategy to enhance the electronic conductivity of active materials and facilitate charge transfer during redox processes \cite{Pati_JPS_24}. Therefore, an {\it in-situ} carbon coating approach is attempted using citric acid, serving simultaneously as a chelating agent and a carbon source. However, this approach failed to yield a pure-phase NNCP sample (see Fig.~S7 of \cite{SI}). Consequently, an ex-situ carbon coating strategy is adopted where first, the bare NNCP powder was mixed with 10 wt\% sucrose as the carbon precursor and heat-treated at 750$\degree$C for 6 hr under an Ar atmosphere. Although, the obtained product appeared black, the XRD analysis revealed the presence of additional impurity phases alongside the main NNCP phase (see Fig.~S7 of \cite{SI}). To overcome this issue, the bare NNCP was instead combined with 10 wt\% acetylene black and subjected to the heat treatment (600$\degree$C for 6 hr in Ar). The resulting NNCP-Air/AB composite exhibited a phase-pure structure, as confirmed by the XRD pattern [see Fig.~\ref{Strc}(c)], without any detectable secondary phases. The NNCP-Air/AB composite is subsequently evaluated electrochemically, with the upper cut-off voltage deliberately reduced to investigate whether the irreversibility observed under high-voltage operation could be verified. From the EIS measurement presented in Fig.~\ref{CV_GCD}(a3), a significant reduction in the resistance values of the NNCP-Air/AB material is quite evident, where the observed values are R$_{SEI}$ = 192.5 $\Omega$,  R$_{ct1}$ = 240.9 $\Omega$, and R$_{ct2}$ = 1053 $\Omega$. A significant drop in the resistance value of the primary charge transfer resistance is observed, as expected from the carbon coating of the bare sample. Further, the CV results in Fig.~\ref{CV_GCD}(b3) show that electrochemical activity at 3.9 V and 4.5 V remains the same in the NNCP-Air/AB cathode. Even when the upper cutoff voltage was reduced to 4.7 V, no significant improvement in the reversibility of the material is observed. The GCD profile in Fig.~\ref{CV_GCD}(c3) showed that a charge capacity of 46.8 mAh/g is achieved but, no plateau or activity is found in the discharging curve, and a discharge capacity of only 5 mAh/g can be achieved, with no change in subsequent cycles (see Figs.~S8(a, b) of \cite{SI}). The discharge capacity in this case is more than the NNCP-Air and NNCP-Ar cases; however, it can find its origin from the charge storage capacities of the carbon coating. Note that the carbon coating can be an effective strategy for improving the electronic conductivity of poorly conducting cathodes. It should be emphasized that it primarily enhances surface conduction, while the bulk conductivity remains intrinsic to the material \cite{Sharma_CCR_25, Wang_EES_12}. 

Finally, to evaluate the structural stability, the XRD pattern of the NNCP-Ar electrode is analyzed after 5 cycles. The diffraction peaks remain consistent with those of the pristine sample, confirming that the NNCP retains its crystallinity even after cycling at higher voltages [see Figs.~\ref{Cycled}(a, b)]. The PO$_4$ tetrahedra also remain largely unaffected, with only a minor shift in the average P--O bond length from 1.536 \AA\ to 1.555 \AA. In contrast, a slight contraction is observed in the average Ni/Cr--O bond length, decreasing from 2.012 \AA\ to 1.933 \AA. Examining the sodium-ion sublattice, the Na1 and Na2 occupy sixfold (octahedral) and eightfold (distorted dodecahedral) oxygen coordination environments, respectively. The Na2 site shows only a marginal increase in the average Na2--O bond length (2.586 \AA\ to 2.600 \AA), accompanied by a small reduction in its polyhedral volume (26.023 \AA$^3$ to 25.688 \AA$^3$). However, the Na1 site undergoes a more significant change, where the average Na1--O bond length increases from 2.412 \AA\ to 2.588 \AA, and the corresponding Na1O$_6$ octahedral volume expands from 15.933 \AA$^3$ to 18.200 \AA$^3$. This indicates that sodium ions at the Na1 sites become weakly bonded to the surrounding oxygen ions. These structural variations, particularly the pronounced changes at the Na1 site, are expected to influence the energy landscape for sodium-ion migration strongly and may impact the overall electrochemical performance. The calculated sodium-ion migration energy profiles for the pristine and cycled electrodes are shown in Figs.~\ref{Cycled}(c, d). In the pristine sample [see Fig.~\ref{Cycled}(c)], the Na1 represents the most stable site, while Na2 serves as the accessible site for sodium de/intercalation, and s1 are the sites where sodium ions are least stable. The sodium transport proceeds along the Na1--Na2--Na1 pathway, encountering a migration barrier of 0.468 eV. However, after cycling [see Fig.~\ref{Cycled}(d)], the energy landscape is altered, the migration barrier decreases to 0.324 eV, and the Na2 becomes more stable than Na1. The redistribution of site stability, together with the observed bond length variations and polyhedral volume changes, suggests that  structural distortions at the sodium sites strongly influence the Na$^+$ migration mechanism during electrochemical cycling. Nevertheless, the calculated energy barriers remain within a feasible range for ion transport, indicating that ionic mobility alone cannot account for the observed electrochemical irreversibility. Here, we speculate that the inability of Ni to stabilize a hole-polaron state, combined with the time-dependent degradation commonly reported for Cr-based NASICON cathodes, may hinder the smooth nucleation of the discharged phase \cite{Kundu_CM_15, Johannes_PRB_12, Kawai_ACSAEM_18, Herklotz_EA_14}. Nevertheless, a more detailed investigation is required to pinpoint the origin of this irreversibility. 
 
\section{\noindent ~Conclusion}

In summary, the Na$_4$NiCr(PO$_4$)$_3$ was explored as a promising high-voltage NASICON-type cathode for sodium-ion batteries. Three samples, namely NNCP-Air, NNCP-Ar, and NNCP-Air/AB, were synthesized, then performed physical characterization and finally electrochemically tested. Although, the materials exhibited promising charge capacities around 4.5 V, no discharge capacity was observed, indicating the absence of reversible sodium-ion intercalation. The observed activity at 4.5 V is possibly associated with the Cr$^{3+}$/Cr$^{4+}$ redox couple, while both Ni redox processes could occur, possibly beyond 5.0 V range. Structural analysis and bond valence energy landscape calculations confirmed well-connected Na$^+$ diffusion pathways with a moderate migration barrier of 0.468 eV, indicating that ionic transport is not the primary factor behind the poor reversibility. It is speculated that the poor electronic conduction or redox limitations are inherent to the Ni-Cr polyanionic framework. These results highlight the challenges of harnessing the high operating voltage of Ni- and Cr-based redox couples in NASICON hosts, emphasizing the need for targeted strategies, such as doping, structural modifications, and electrolyte optimization, to unlock the practical potential of high-voltage cathodes with multi-electron reactions in SIBs.

\section{\noindent ~Acknowledgments}

MS and PS thanks the CSIR-HRDG and UGC, respectively, for the fellowship support. RSD acknowledges the DST for financially supporting the research facilities for sodium-ion batteries through {\textquotedblleft}DST-IIT Delhi Energy Storage Platform on Batteries" (project no. DST/TMD/MECSP/2K17/07) and from SERB-DST (now ANRF) through a core research grant (file no.: CRG/2020/003436).  We thank IIT Delhi for providing research facilities for sample characterization (the XRD and Raman at the physics department; as well as the FE-SEM, EDS, and XPS at CRF). 

%\section{\noindent ~Conflict of interest }
%The authors declare that there are no conflicts of interest associated with this article. 

%\section{\noindent ~Data Availability }
%The data that support the findings of this study are available upon reasonable request.

\end{document}